\documentclass[conference,review,anonymous]{IEEEtran}
\IEEEoverridecommandlockouts
\usepackage{siunitx}
\def\BibTeX{{\rm B\kern-.05em{\sc i\kern-.025em b}\kern-.08emT\kern-.1667em\lower.7ex\hbox{E}\kern-.125emX}}
    
\usepackage[belowskip=-10pt,aboveskip=2pt, font=small]{caption}
\usepackage[labelformat=simple,skip=0pt]{subcaption}
\usepackage[compress]{cite}
\usepackage{amsmath,amssymb,amsfonts}
\usepackage{algorithmic, algorithm}
\usepackage{graphicx}
\usepackage{textcomp}
\usepackage{multirow}
\usepackage{multicol}
\usepackage{url}
\usepackage{xcolor}
\usepackage{comment}
\usepackage{makecell}
\usepackage[normalem]{ulem}

\makeatletter
\newcommand\notsotiny{\@setfontsize\notsotiny\@vipt\@viipt}
\makeatother

\usepackage{listings}
\definecolor{codered}{rgb}{0.75,0.0,0.0}
\definecolor{codegreen}{rgb}{0.0,0.4,0.0}
\definecolor{codeblue}{rgb}{0.0,0.0,0.6}
\definecolor{light-gray}{gray}{0.95}

\newcommand{\ignore}[1]{}
\newif\ifsubmit
\submitfalse
\ifsubmit
    \newcommand{\mert}[1]{}
    \newcommand{\vikram}[1]{}
    \newcommand{\carl}[1]{}
    \newcommand{\wenmei}[1]{}
    \newcommand{\RN}[1]{}
\else
    \definecolor{gray}{rgb}{0.66, 0.66, 0.66}
    \definecolor{dgreen}{rgb}{0.00, 0.75, 0.00}
    \definecolor{dblue}{rgb}{0.00, 0.00, 0.75}
    \newcommand{\mert}[1]{[{\color{dgreen}MH: #1}]}
    \newcommand{\vikram}[1]{[{\color{red}VK: #1}]}
    \newcommand{\carl}[1]{{\color{red}CP: #1}}
    \newcommand{\wenmei}[1]{[{\color{dgreen}ASSIGN: #1}]}
    \newcommand{\RN}[1]{[{\color{blue}RN: #1}]}
    
\fi

\author{\IEEEauthorblockN{Mert Hidayeto\u{g}lu, Carl Pearson, Vikram Sharma Mailthody, Eiman Ebrahimi\IEEEauthorrefmark{3},\\ Jinjun Xiong\IEEEauthorrefmark{4}, Rakesh Nagi, and Wen-mei Hwu}

\IEEEauthorblockA{University of Illinois at Urbana-Champaign, \IEEEauthorrefmark{3}NVIDIA, \IEEEauthorrefmark{4}IBM Research}

\IEEEauthorblockA{
\{hidayet2, pearson, vsm2\}@illinois.edu, eebrahimi@nvidia.com,  jinjun@us.ibm.com, \{nagi,w-hwu\}@illinois.edu}}
\begin{document}

% \title{At-Scale Inference of Sparse Deep Neural Network on GPUs\vspace{-1ex}}

\title{At-Scale Sparse Deep Neural Network Inference With Efficient GPU Implementation}

% \title{Efficient Inference on GPUs for the Sparse Deep Neural Network\vspace{-1ex}}

\maketitle
\thispagestyle{plain}
\pagestyle{plain}

%% Author information
%% Contents and number of authors suppressed with 'anonymous'.
%% Each author should be introduced by \author, followed by
%% \authornote (optional), \orcid (optional), \affiliation, and
%% \email.
%% An author may have multiple affiliations and/or emails; repeat the
%% appropriate command.
%% Many elements are not rendered, but should be provided for metadata
%% extraction tools.

% \iffalse%old ACM author format, change to IEEE after accepted.
% \author{Mert Hidayeto\u{g}lu${}^1$, Carl Pearson${}^1$, Vikram Sharma Mailthody${}^1$, Rakesh Nagi${}^1$, Jinjun Xiong${}^2$ and Wen-mei W.~Hwu${}^1$}
% \affiliation{${}^1$University of Illinois at Urbana-Champaign, IL 61810, USA}%, ${}^2$Argonne National Laboratory, ${}^3$College of William \& Mary, ${}^4$Northwestern University, and ${}^5$University of Chicago}
% \affiliation{${}^2$ IBM Thomas J. Watson Research Center, Yorktown Heights, NY, 10598, USA}
% \affiliation{${}^3$College of William \& Mary, VA 23185, USA}

%\email{hidayet2@illinois.edu}

% \renewcommand{\shortauthors}{M. Hidayeto\u{g}lu et al.}
 
% \fi

\begin{abstract}
This paper presents GPU performance optimization and scaling results for inference models of the Sparse Deep Neural Network Challenge 2020.
Demands for network quality have increased rapidly, pushing the size and thus the memory requirements of many neural networks beyond the capacity of available accelerators.
Sparse deep neural networks (SpDNN) have shown promise for reining in the memory footprint of large neural networks.
However, there is room for improvement in implementing SpDNN operations on GPUs.
This work presents optimized sparse matrix multiplication kernels fused with the ReLU function.
The optimized kernels reuse input feature maps from the shared memory and sparse weights from registers.
For multi-GPU parallelism, our SpDNN implementation duplicates weights and statically partition the feature maps across GPUs.
Results for the challenge benchmarks show that the proposed kernel design and multi-GPU parallelization 
achieve up to 180 TeraEdges per second inference throughput.
These results are up to 4.3$\times$ faster for a single GPU and 
an order of magnitude faster at full scale than those of the champion of the 2019 Sparse Deep Neural Network Graph Challenge for the same generation of NVIDIA V100 GPUs.  
Using the same implementation\footnote{Our code is open-source at
\url{https://github.com/merthidayetoglu/SpDNN_Challenge2020}}, we also show single-GPU throughput on NVIDIA A100 is 2.37$\times$ faster than V100.

\end{abstract}

%As the use of neural networks becomes pervasive in many domains, the accuracy requirement has also increased rapidly, pushing the size of many neural networks in terms of number of weights beyond the memory capacity of available accelerators.

%The challenge provides a performance testbed involving 12 synthetic SpDNNs with various numbers of layers and neurons per layer. The layers represent pruned  (sparse) weights followed by biased ReLU function. 
%\RN{less about the challenge and more about the contributions.}

\section{Introduction}

Deep learning (DL) has seen great progress over the last decade and demonstrated substantial accuracy improvement over a range of machine learning tasks such as image classification~\cite{ameobanet}, object recognition~\cite{zoph2020rethinking}, language modeling~\cite{megatronlm,gpt3}, and language translation~\cite{googlenmt}.
% These large accuracy improvements can be attributed to their ever increasing size of deep learning models. 
Scaling up DL models along with increased amount of training data have proven to be an effective approach to improve the model accuracy~\cite{gpt3,megatronlm,huang2018gpipe}.
The progress of DL applications involves not only the algorithmic improvements, but also the unprecedented computational throughput provided by GPUs.
%\RN{Some remark about the advances in hardware that have made this feasible from a computational perspective.}\mert{addressed}

On several machine learning tasks, 
%As the neural networks becomes prominent algorithm to address challenging problems, 
%As neural networks are applied to more problems, 
the size of state-of-the-art deep neural networks (DNN) has grown beyond the memory limits of available accelerators~\cite{megatronlm,huang2018gpipe}.
To address this, the DL community has taken an algorithmic approach of sparsifying the DNN using techniques such as pruning~\cite{han2015deepcompression,radixnet,efficientnet}. 
These optimization convert a dense DNN into a sparse DNN.
%
%However, sparse DNNs present implementation challenge for GPUs, which are designed for dense and regular computations. 

Sparse DNNs present unique scalablity challenges.
Realizing this, in 2019 MIT/IEEE/Amazon proposed the Sparse DNN Challenge as an extension to the Graph Challenge~\cite{kepner2020graphchallengeorg,kepner2019,graphchallenge18,graphchallengetc19,graphchallengektruss19}. %,graphchallenge18,graphchallengetc19,graphchallengektruss19}. to add in camera ready
The Sparse DNN Challenge is created by leveraging the collective knowledge of machine learning, high-performance computing and graph analytics communities on emerging AI systems. 
The objective of the Sparse DNN Challenge is to build scalable algorithms and systems for sparse AI analytics. %\RN{Language.}

The Sparse DNN Challenge provides model structure, trained model weights, and inputs as a foundation for comparison of inference performance.
Several models with varying number of layers (120 to 1920 layers) and neuron connections (4 Million to 4 Billion) are posted in the challenge. 
Such a large variation allows system and algorithm designers to evaluate efficient, scalable implementations against a variety of network types.

In this work, we propose efficient sparse algorithms and demonstrate performant kernel implementations for sparse DNN inference on GPUs. 
GPUs have become a de-facto accelerator for DNN computation. 
However, traditionally GPUs are designed for \textit{dense} DNN operations and their performance is considerably reduced for \textit{sparse} DNN computations. 
Using a baseline sparse DNN implementation, we identify the causes of performance bottlenecks. 
In our baseline implementation, we store the weights as sparse matrices in the compressed sparse row (CSR) storage format while the inputs are represented as dense matrices in the column-major data layout in memory. 
We observe that irregular memory accesses to the input matrix and redundant accesses to the weight matrices are the  primary causes for inefficiency in the naive sparse DNN implementation. 

To address this, we propose a few optimizations, namely, register tiling, shared-memory tiling, and compact index representation to create optimized kernel implementation fused with ReLU activation for sparse DNN inference. 
In the optimized fused kernel, the weights are stored in sliced ELLPACK format for efficient memory access while the inputs are retained in the column-major layout in memory.
Together these optimizations minimize the total number of irregular memory access and redundant access to the global memory, thus providing significant performance improvements. 
Although variants of these optimizations have been applied in sparse matrix mutiplication (SpMM) kernels~\cite{mertpaper}~\cite{mertpaper2}, in this work we tune them for sparse DNN computation. 

Our evaluation shows, 
compared to the baseline fused kernel implementation, the optimized fused kernel provides up to 11.84$\times$ speedup.
Overall, the proposed optimizations achieve up to 14.30 TeraEdges, i.e., 10${}^\textrm{12}$ edges, per second performance on a single V100 GPU. 
Compared with prior champions, our single GPU implementation on the same generation of NVIDIA Volta V100 GPU is up to 4.3$\times$ faster. 
Compared to an implementation based off of NVIDIA's cuSPARSE, our proposed kernels can provide 200$\times$ speedup.
In addition, we evaluate the performance of our optimized fused kernel on the latest generation NVIDIA Ampere A100 GPU, and show that out-of-the-box execution of our optimized kernel achieves up to 20.99 TeraEdges per second (2.37$\times$ faster than V100 GPU). 

% We propose a load-balanced batch parallelization to improve scaling on GPU clusters.
To measure the algorithm scalablity, we perform strong scaling using batch parallelism and scale up to 768 V100 GPUs on department of energy's Summit supercomputer at Oak Ridge National Laboratory \cite{summit}. 
On a single node, when the number of V100 GPUs are increased from one to six, optimized fused kernel achieves up to 5.37$\times$ speedup, and thus providing a scaling efficiency of 89.5\%.
Our evaluation shows that, our proposed implementation can provide up to 51.8$\times$ speed up when the number of GPUs is increased to 768 (or 128 nodes with 6 GPUs in each node) compared to the single V100 GPU implementation. 
Compared to 2019 Sparse DNN Champions, our scale-out implementation is 3.25--19.13$\times$ faster, providing up to 180 TeraEdges per second. % and \RN{where?} scalability is limited by the number of inputs.

Overall, we make the following main contributions: %\RN{Rethink?}
\begin{enumerate}
    \item We present novel fused SpMM kernel design that is optimized for providing data reuse during DNN inference.
    \item We perform extensive at-scale benchmarking on up to 768 V100 GPUs as well as the new A100 GPUs for the Sparse Deep Neural Network Challenge dataset.
\end{enumerate}
\section{Background \& Motivation}
\label{sec:backmotiv}

In this section, we provide a brief overview of the Sparse DNN Challenge and discuss the limitations in naive sparse DNN inference implementation.

% \begin{figure}[h!]
%     \centering
%     \includegraphics[width=8.5cm]{Figures/layer.pdf}
%     \caption{Comparison of Dense and Sparse DNN neuron connection}
%     \label{fig:layer}
% \end{figure}

\subsection{Overview of Sparse DNN Challenge}
The {Sparse DNN Challenge} specifies
%is the latest challenge in the MIT/IEEE/Amazon Graph Challenge contest that performs 
%requires the participants to perform DNN inference computation on 
a collection of large sparse DNNs models~\cite{kepner2020graphchallengeorg,kepner2019} %,graphchallenge18,graphchallengetc19,graphchallengektruss19}. to add in camera ready. 
%These sparse DNNs model
that are representative of the latest trends in %the deep learning community to 
addressing challenging machine learning tasks. % \RN{and inference} - ML covers inference and training. So not needed.
The challenge provides model structure (number of layers and size of layer), and model weights for computing sparse DNN inference on a given input dataset.

\subsubsection{Formulation {of} Sparse Layer}
{ For a sparse DNN with $L$ layers, each sparse layer computation can be formulated as}
\begin{equation}
\label{eq:activation}
    \boldsymbol{Y}_{l+1}={\textrm{ReLU}}(\boldsymbol{W}_l\times\boldsymbol{Y}_l+\boldsymbol{B})
\end{equation}
where $l$ is the layer number, 
$\boldsymbol{Y}_l$ and $\boldsymbol{Y}_{l+1}$ are $N\times M$ matrices of $M$ input and output features of length $N$, respectively, stored in column-major format,
$\boldsymbol{W}_l$ is an $N \times N$ matrix of activation weights,
$\boldsymbol{B}_l$ is an $N \times M$ bias matrix for each output,
and ReLU is the activation function. Here, ReLU activation function promotes sparsity and is defined as %$\textrm{ReLU}\{x\} = \textrm{max}(0,\textrm{min}(x,32))$. 
% however, it can be generalized to other activation functions, i.e., sigmoid function.
$\textrm{ReLU}(x) = \textrm{max}\{0,\textrm{min}\{x,32\}\}$.
The weight matrix represents a general pattern of neuron connections:
for a fully-connected layer, $\boldsymbol{W}_l$ would be entirely non-zero,
for a convolution layer, $\boldsymbol{W}_l$ would be banded, and for a general sparse layer, $\boldsymbol{W}_l$ is a general sparse matrix.

\subsubsection{Steps Involved {in} Sparse DNN Challenge}
Algorithm~\ref{alg:psuedosparsednn}  describes the  high-level steps involved in computing sparse DNN inference in the {Sparse DNN Challenge}~\cite{kepner2020graphchallengeorg,bisson2019gpu}. 
The challenge provides datasets comprising of input data for the neural network, weights for each layer in the network, bias values for each layer and finaly the ground truth to validate if the results are correct while computing inference. 

\setlength{\textfloatsep}{8pt}% Remove \textfloatsep
\begin{algorithm}
  \caption{Outline of Sparse DNN Challenge Algorithm}
  \begin{algorithmic}[1]
    \STATE  Read input $\boldsymbol{Y}_0$ and model weights ($\boldsymbol{W}_0$,$\boldsymbol{W}_1$ ... $\boldsymbol{W}_{L-1}$) from binary files
    \STATE  Initialize bias vector $\boldsymbol{B}$ with a constant
    \STATE  Evaluate Equation~(\ref{eq:activation}) for all the layers, starting from $l=0$ and until $l=L-1$
    \STATE Use $\boldsymbol{Y}_{L-1}$ to determine categories and compare with the ground truth. 
    \STATE Measure and report performance 
  \end{algorithmic}
  \label{alg:psuedosparsednn}
\end{algorithm}

The input to the neural network is an interpolated sparse version derived from the MNIST dataset and consists of 60,000 images stored in TSV format. 
Each of the MNIST 28$\times$28 pixel images are resized to 32$\times$32 (1024 neurons), 64$\times$64 (4096 neurons), 128$\times$128 (16384 neurons) and 256$\times$256 (65536 neurons) pixels. 
All resized  images are thresholded to ensure values reside between 0 to 1. 
Then, they are linearized and stacked together to  obtain  60k×1K,  60k×4K,  60k×16K,  and  60k×64K  input feature matrices.
This allows to store each image in a single row in the TSV while each column representing a non-zero pixel location with a value of 1. 
The challenge also provides three different sparse DNN models comprising of 120, 480, and 1920 layers. Thus effectively  creates a total of 12 DNNs (\{1024, 4096, 16384, 65536\} neurons $\times$ \{120, 480, 1920\} layers), {publicly available~\cite{hpec}.}

Since such large sparse DNN networks are not publicly available, the challenge generates the weight matrix using RadiX-Net synthetic sparse DNN generator~\cite{radixnet}. 
The synthetic generator is capable of creating pre-determined DNNs with 32 connections per neuron and ensuring there exists an equal number of paths between inputs and intermediate  layers. 
All neurons have weights and biases set to 1/16. However, our kernel design and implementation could work on real-life data with any arbitrary bias values and any sparse layer with an arbitrary sparsity pattern.

\subsection{The Baseline GPU Implementation}\label{sec:naive}
\begin{figure}[h]
    \centering
    \includegraphics[width=8.5cm]{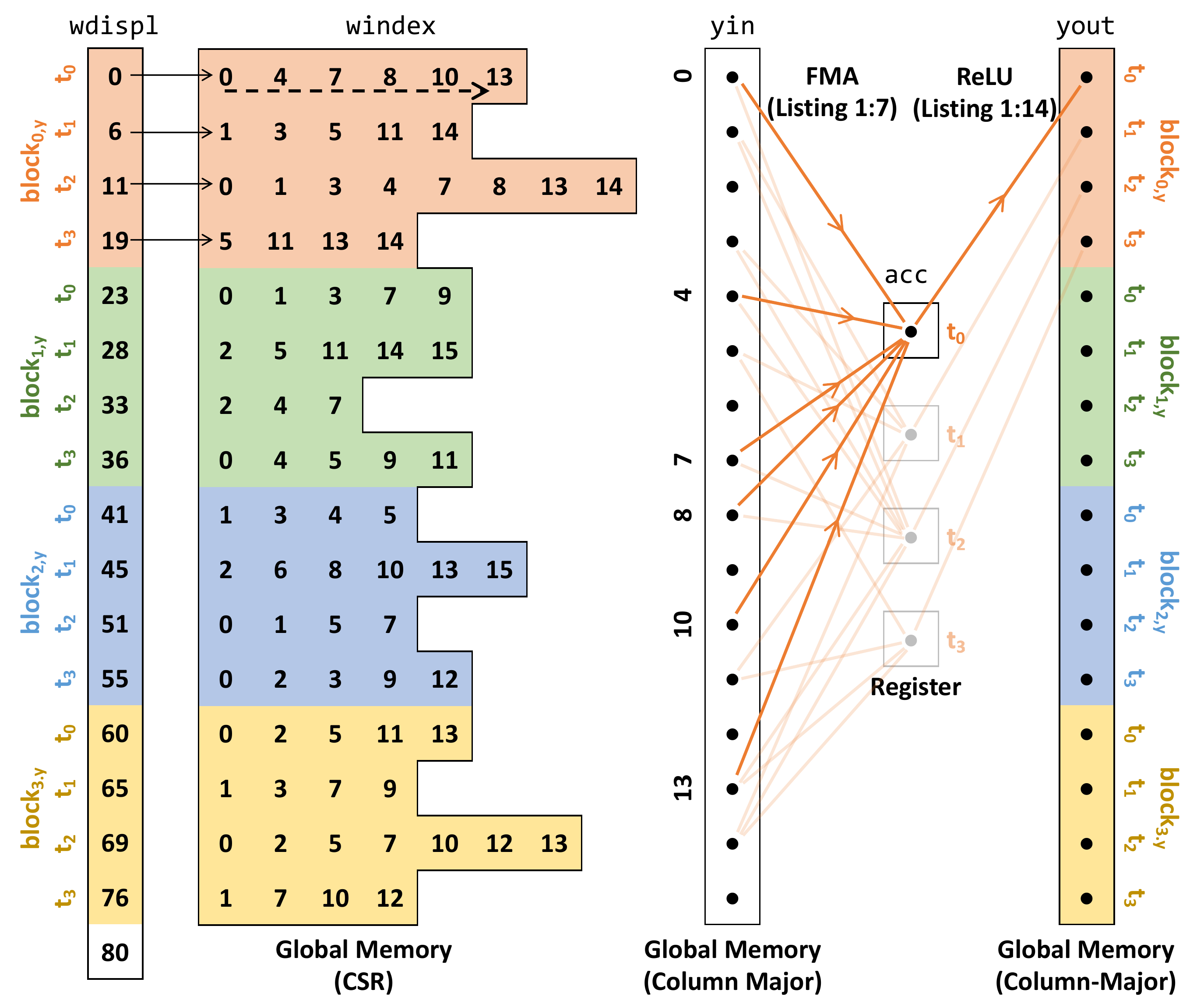}
    \vspace{1mm}
    \caption{Baseline fused kernel implementation. Each thread irregularly accesses the input features and accumulates the output in register.}
    \vspace{4mm}
    \label{fig:naive_denseXsparse}
\end{figure}

We present a baseline implementation of sparse DNN using GPUs and discuss the bottlenecks to motivate our optimizations.
%We picked GPUs as they have become the de facto accelerator for DNN computation.
%We identify the computational inefficiencies of a naive GPU implementation by investigating 
The baseline implementation consists a sparse$\times$dense SpMM kernel fused with ReLU function for GPU execution. 
In this implementation, shown in Listing~\ref{list:naive_sparseXdense}, each thread produces a single output element in $\boldsymbol{Y}_{l+1}$, and gathers data from the corresponding column of $\boldsymbol{Y_l}$. 
 Each thread reads a row of $\texttt{windex}$ (column indices of non-zero weight elements in the CSR format)  and $\texttt{wvalue}$ (the value of the non-zero weight elements) from sparse $\boldsymbol{W}_l$ matrix stored in CSR format, as depicted in Figure~\ref{fig:naive_denseXsparse}. 
 Note that $\texttt{windex}$ and
$\texttt{wvalue}$ have the same layout and access patterns so we omitted $\texttt{wvalue}$ from Figure~\ref{fig:naive_denseXsparse}. Also, each $\texttt{wdispl}$ element gives the starting location of its corresponding row of non-zero elements in the CSR format. That is, all $\texttt{windex}$ elements in Figure~\ref{fig:naive_denseXsparse} are stored in a linear array with the starting point of each row of non-zero elements delineated by the $\texttt{wdispl}$ elements.

\begin{lstlisting}[language=C++, caption={Baseline Implementation
\label{list:naive_sparseXdense}},basicstyle=\notsotiny\ttfamily,tabsize=1,literate={\ \ }{{\ }}1,numbers=left,xleftmargin=6ex, frame=bt]
__device__ float __ReLU(float x){
   return x<0.0?0.0:x>32.0?32.0:x;
};
__global__ void fused_ReLU(float *yout, float *yin, int neuron,
  int *wdispl, int *windex, float *wvalue, float *bias, int *active, 
  int *category){
    int xoff = blockIdx.x*blockDim.x+threadIdx.x;
    int yoff = category[blockIdx.y]*neuron;
    float acc = 0;
    for(int n = wdispl[xoff]; n < wdispl[xoff+1]; n++)
        acc += yin[yoff+windex[n]]*wvalue[n];
    acc = __ReLU(acc+bias[xoff]);
    if(acc > 0){
        yout[blockIdx.y*neuron+xoff] = acc;
        atomicAdd(active+blockIdx.y,1);
    }
}

//INFERENCE LOOP AT HOST CPU
for(int l = 0; l < layer; l++){
    dim3 grid(neuron/blocksize,mybatch);
    dim3 block(blocksize);
    cudaMemset(active_d,0,sizeof(int)*mybatch);
    fused_ReLU<<<grid,block>>>(nextfeat_d,currfeat_d,neuron,
      wdispl_d[l],wind_d[l],wval_d[l],bias_d,active_d,categories_d);
    cudaMemcpy(active,active_d,sizeof(int)*mybatch,D2H);
    int feature = 0;
    for(int k = 0; k < mybatch; k++){
        if(active[k]){
            globalcategories[feature] = globalcategories[k];
            categories[feature] = k;
            feature++;
        }
    }
    cudaMemcpy(categories_d,categories,sizeof(int)*feature,H2D);
    mybatch = feature;
    FEATPREC *tempfeat_d = currfeat_d;
    currfeat_d = nextfeat_d;
    nextfeat_d = tempfeat_d;
}
\end{lstlisting}

For simplicity in the figure, we assume a toy example where each block consists of four threads and the warp size is two. %\RN{Warp size of two, OR "with" a warp size of two}
As a result, the GPU kernel deploys $N/B$ thread blocks for each feature, where $N$ is the number of output feature elements (pixels), and $B$ is the block size. 
Here, each block accesses a portion of $\boldsymbol{W}_l$. 
As each thread accesses the input, it performs a fused multiply-and-add (FMA) operation with the corresponding activation weight, accumulating the result in its register \texttt{acc}. 
Finally, each thread adds the bias and activate (or deactivate) the output, and writes the output to global memory.

While the baseline implementation exhibits a high degree of thread-level parallelism for large output features, it is inefficient for several reasons. 
First, the global memory accesses to the input matrix are not only uncoalesced, but also irregular: Figure~\ref{fig:naive_denseXsparse} highlights the memory accesses of the first thread of the first block. An iregular subset of the $\boldsymbol{Y}_l$ elements are read by each thread. 
Second, threads that generate equivalent elements in different output maps redundantly reads the same row of the $\boldsymbol{W}_l$ matrix. 
As a result the weight matrix is read $M$ times. 
This results in wasting memory bandwidth that could have been used for the needed computation. 
Third, the load imbalance among threads caused by different number of non-zero elements in adjacent rows of $\boldsymbol{W}_l$ yields thread divergence within warps. 
We address these limitation in the next section.

\section{Proposed Algorithm Design}
\label{sec:singlegpu}

In this section we discuss in detail the 
%four performance and one memory optimizations 
three GPU performance optimizations {to improve the inference rate of the baseline implementation, and two memory optimizations to fit large models into limited GPU memory.} 
These optimizations involve register tiling, shared memory tiling, efficient access to weight matrix $\boldsymbol{W}_l$, compact index representation and out-of-core storage of weight matrix $\boldsymbol{W}_l$. 
Listing~\ref{listing:optimized_denseXsparse} shows the optimized-fused kernel implementation for sparse DNN computation. Next we will describe each of these optimizations {and tuning parameters} in detail.

\subsection{GPU Kernel Optimizations}

\subsubsection{Register Tiling}%\vikram{Ready for review}
The most obvious inefficiency of the baseline fused kernel (shown in Listing~\ref{list:naive_sparseXdense}, line 11) is the duplicated access to the weight matrix across output features.  
To provide reuse of the weight matrix $\boldsymbol{W}_l$ from register, the optimized kernel groups (\texttt{MINIBATCH}) multiple active features and performs the ReLU computation for all of them in a single step.
This reuses the index (\texttt{windex}) and value (\texttt{wvalue}) elements in each thread for \texttt{MINIBATCH} features as shown in Listing~\ref{listing:optimized_denseXsparse}, lines 23 and 24, where \texttt{MINIBATCH} is the number of features in each minibatch.

When \texttt{MINIBATCH} is increased, the register data reuse improves, thus increasing the arithmetic intensity. This saves memory bandwidth for reading the weight matrices, but also increases register useage.
A \texttt{MINIBATCH} value of $12$ is selected for optimal performance, which balances increased reuse with memory spills from increased register pressure.

\subsubsection{Shared-Memory Tiling}
We now address the duplicate and irregular accesses to input features.
Each input feature element is potentially accessed multiple times %by different neurons, 
by multiple threads
and they are accessed in an irregular patterns controlled by the weight sparsity pattern expressed by \texttt{windex}.
This wastes memory bandwidth and increases access latency.
For example, for $block_{0,y}$ in Figure~\ref{fig:naive_denseXsparse}, both $t_0$ and $t_2$ access \texttt{yin[4]} but at different iterations for their execution.

\begin{lstlisting}[language=C++, caption={Optimized Kernel Implementation
\label{listing:optimized_denseXsparse}},basicstyle=\notsotiny\ttfamily,tabsize=1,literate={\ \ }{{\ }}1,numbers=left,xleftmargin=6ex, frame=bt]
__global__ void opt_ReLU(float *yout, float *yin, int neuron,
  int *buffdispl, int *mapdispl, unsigned short *map,
  int *wdispl, unsigned short *windex, float *wvalue,
  float *bias, int *active, int *category){
    __shared__ float shared[BUFFSIZE];
    int wind = threadIdx.x%WARPSIZE;
    int yoff = blockIdx.y*MINIBATCH;
    int xoff = blockIdx.x*blockDim.x+threadIdx.x;
    float acc[MINIBATCH] = {0.0};
    for(int buff = buffdispl[blockIdx.x]; 
            buff < buffdispl[blockIdx.x+1]; buff++){
        int mapnz = mapdispl[buff+1]-mapdispl[buff];
        for(int n = threadIdx.x; n < mapnz; n += blockDim.x){
            int ind = map[mapdispl[buff]+n];
            for(int f = 0; f < MINIBATCH; f++){
                shared[f*buffsize+n] = yin[category[yoff+f]*neuron+ind];
        }
        __syncthreads();
        int warp = (buff*blockDim.x+threadIdx.x)/WARPSIZE;
        for(int m = wdispl[warp]; m < wdispl[warp+1]; m++){
            int ind = windex[m*WARPSIZE+wind]; 
            float val = wvalue[m*WARPSIZE+wind]; 
            for(int f = 0; f < MINIBATCH; f++)
                acc[f] += shared[f*buffsize+ind]*val;
        }
        __syncthreads();
    }
    for(int f = 0; f < MINIBATCH; f++){
        acc[f] = __ReLU(acc[f]+bias[xoff]));
        if(acc[f] > 0){
            yout[(yoff+f)*neuron+xoff] = acc[f];
            atomicAdd(active+blockIdx.y*MINIBATCH+f,1);
        }
    }
}
\end{lstlisting}

\begin{figure}[h]
    \centering
    \includegraphics[width=\columnwidth]{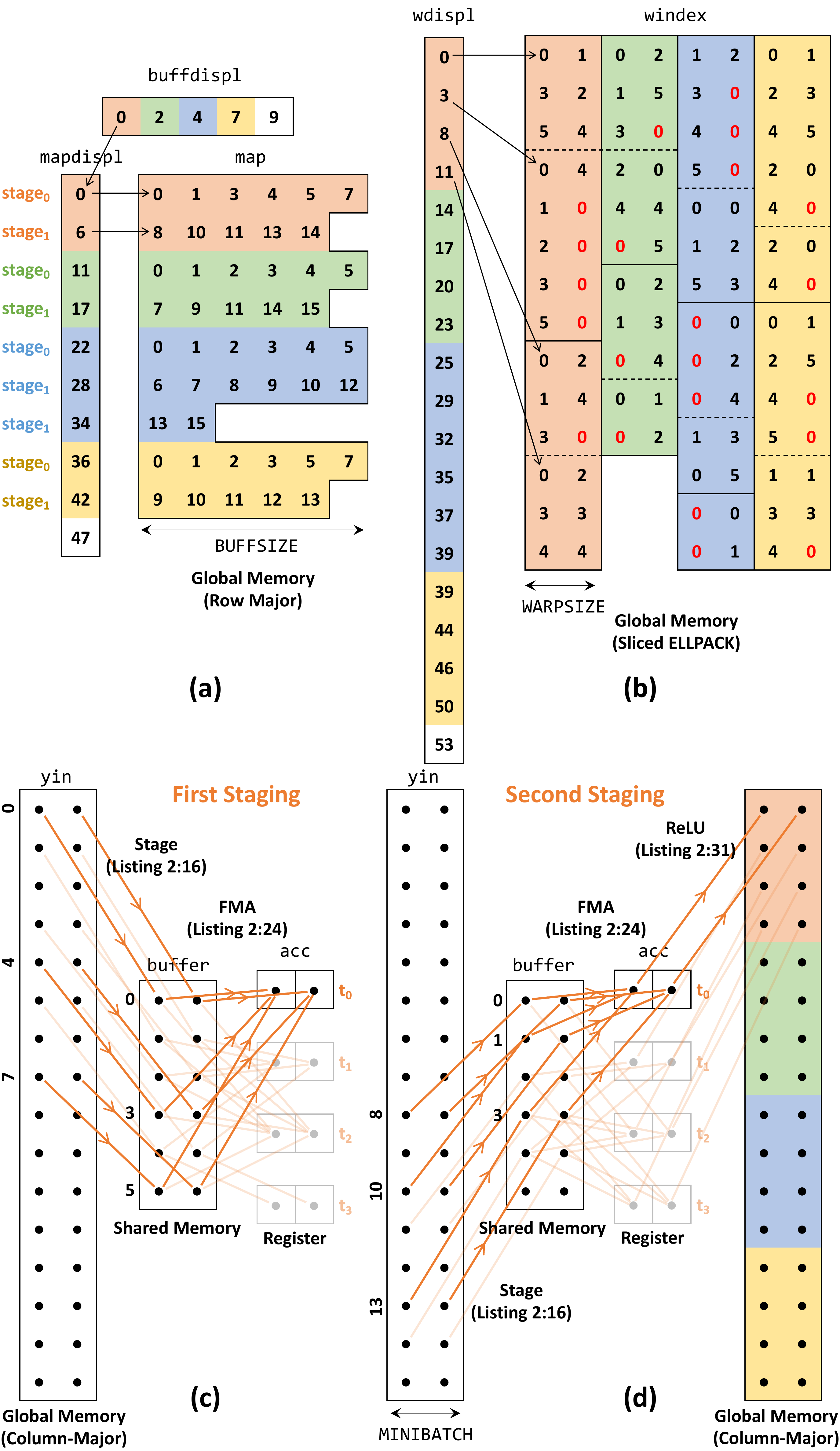}
    % \vspace{-2ex}
    \caption{Optimized-fused kernel execution example.}% {\color{red} WMH: label the Figure as (a) upper left, (b) upper right, (c) lower left and (d) lower right.}
    \vspace{2mm}
    \label{fig:optimized_denseXsparse}
\end{figure}

\begin{table*}
\centering
% \captionsetup{font=large}
\vspace{1cm}
\caption{Inference Throughput (TeraEdges/Second)}
\label{tab:summit}
\renewcommand{\arraystretch}{1.5}
\resizebox{0.95\textwidth}{!}{%
\begin{tabular}{cccccccccccccc}
%\Xhline{0.75pt}
\multicolumn{4}{c}{} & \multicolumn{9}{c}{\textbf{Number of V100 GPUs (Six per Node)}} \\ \cline{5-13} 
\textbf{Neurons} & \textbf{Layers} & \multicolumn{1}{c}{\textbf{Single V100}} & \multicolumn{1}{c}{\textbf{Single A100}} & \multicolumn{1}{c}{\textbf{3}} & \multicolumn{1}{c}{\textbf{6}} & \multicolumn{1}{c}{\textbf{12}} & \multicolumn{1}{c}{\textbf{24}} & \multicolumn{1}{c}{\textbf{48}} & \multicolumn{1}{c}{\textbf{96}} & \multicolumn{1}{c}{\textbf{192}} & \multicolumn{1}{c}{\textbf{384}} & \multicolumn{1}{c}{\textbf{768}}\\ \Xhline{0.75pt}
\multirow{3}{*}{\textbf{1024}} & \textbf{120} & 10.51 (0.225s) & 16.74 (1.59$\times$) &  18.92 & 22.46 & 25.52 & 28.52 & 27.77 & \textbf{29.17} & 27.89 & 29.12 & 29.13 \\ \cline{2-13} 
 & \textbf{480} & 12.87 (0.073s) & 20.99 (1.63$\times$) & 21.47 & 24.34 & 26.92 & 28.73 & 28.43 & \textbf{29.30} & 28.80 & 29.10 & 23.06 \\ \cline{2-13} 
 & \textbf{1920} & 14.30 (0.264s) & 20.68 (1.45$\times$) & 22.26 & 24.77 & 27.33 & 28.70 & 28.58 & 28.60 & 28.73 & \textbf{28.83} & 28.83 \\ \Xhline{0.75pt}
\multirow{3}{*}{\textbf{4096}} & \textbf{120} & 9.45 (0.100s) & 14.27 (1.51$\times$) & 20.69 & 31.36 & 47.82 & 62.03 & 70.31 & 75.81 & 79.11 & 81.13 & \textbf{82.20} \\ \cline{2-13} 
 & \textbf{480} & 11.74 (0.322s) & 18.63 (1.59$\times$) & 28.18 & 40.58 & 56.54 & 67.63 & 73.16 & 77.27 & 80.02 & 79.97 & \textbf{82.22} \\ \cline{2-13} 
 & \textbf{1920} & 13.88 (1.08s) & 19.86 (1.43$\times$) & 30.53 & 44.48 & 62.74 & 72.57 & 73.72 & 76.25 & 79.99 & 80.67 & \textbf{82.32} \\ \Xhline{0.75pt}
\multirow{3}{*}{\textbf{16384}} & \textbf{120} & 6.15 (0.614s) & 11.60 (1.89$\times$) & 16.31 & 28.85 & 50.74 & 64.33 & 89.18 & 111.44 & \textbf{146.88} & 114.87 & 111.30 \\ \cline{2-13} 
 & \textbf{480} & 7.45 (2.027s) & 14.31 (1.92$\times$) & 19.82 & 32.88 & 50.83 & 71.45 & 95.78 & 112.61 & 138.62 & 138.30 & \textbf{139.44} \\ \cline{2-13} 
 & \textbf{1920} & 7.84 (7.704s) & 15.27 (1.95$\times$) & 20.86 & 33.62 & 57.08 & 77.73 & 104.83 & 120.63 & 146.11 & 146.30 & \textbf{146.40} \\ \Xhline{0.75pt}
\multirow{3}{*}{\textbf{65536}} & \textbf{120} & 3.47 (4.352s) & 8.15 (2.35$\times$) & 10.90 & 18.77 & 34.20 & 51.14 & 73.67 & 100.72 & 162.19 & 173.25 & \textbf{179.58} \\ \cline{2-13} 
 & \textbf{480} & 3.83 (15.769s) & 9.08 (2.37$\times$) & 12.13 & 20.39 & 37.63 & 56.66 & 75.29 & 108.06 & 166.15 & \textbf{170.26} & 169.30 \\ \cline{2-13} 
 & \textbf{1920} & 3.93 (61.474s) & 9.33 (2.37$\times$) & 12.47 & 20.88 & 38.81 & 58.08 & 77.55 & 112.01 & 167.43 & 170.06 & \textbf{171.37} \\ \Xhline{0.75pt}
\end{tabular}%
}
\vspace{-4mm}
\end{table*}

To address this inefficiency, we stage irregular accesses through shared memory. This is supported by preprocessing the \texttt{windex} rows for each thread block and build a preloading list (\texttt{map}) of all the \texttt{yin} elements collectively accessed by all the threads in the thread block. For example, the preload list for $block_{0,y}$ in Figure~\ref{fig:naive_denseXsparse} would be [0,1,3,4,5,7,8,10,11,13,14]. 

During execution, the thread block will collect these elements into consecutive entries of a \texttt{buffer} array in the shared memory. For example, \texttt{yin[0]} will be loaded into \texttt{buffer[0]}, \texttt{yin[1]} into \texttt{buffer[1]}, \texttt{yin[3]} into \texttt{buffer[2]}, and so on.
The required tiling data structures are constructed once prior to inference and are reused from global memory for computation of all features.

The second part of the preprocessing step is to update the \texttt{windex} so that the stored indices become indicies into the \texttt{buffer}. %This is illustrated in Figure~\ref{fig:optimized_denseXsparse}. 
For example,  \texttt{windex} for $t_{0,y}$ will be updated  from [0, 4, 7, ...] to [0, 3, 5, ...].
During kernel execution, all threads of a thread block cooperatively collect all the needed \texttt{yin} elements into its shared memory. %preload list to collective footprint from the global memory into the shared memory

%The key insight that helps this optimization is that the random access memory bandwidth of shared memory is higher than global memory.

%
When input feature access footprint of a thread block is larger than the shared-memory size, the irregular accesses are performed in multiple stagings, as illustrated in Figure~\ref{fig:optimized_denseXsparse}(a). Assume that the shared memory can only accommodate six \texttt{yin} elements for each thread block, the footprint for $block_{0,y}$ will be divided into two stages.
The footprint of each stage is mapped into the \texttt{buffer} array independently. Thus, for $block_{0,y}$, \texttt{yin[8]}, \texttt{yin[10]}, and \texttt{yin[13]} that are accessed in the second stage are mapped into \texttt{buffer[0], buffer[1], buffer[3]}, as illustrated in Figure~\ref{fig:optimized_denseXsparse}(d).

For each stage, Listing~\ref{listing:optimized_denseXsparse}, line 12--17, shows loading of feature data to shared memory using the \texttt{map} data structure. Then feature data is accessed irregularly from shared memory at line 24 according to the updated address read from \texttt{windex}.
 
%\subsubsection{Thread Coarsening}
%Shared-memory tiling requires all blocks to load all tiles within the corresponding rows, i.e., the input features are read $N/B$ times, where $B$ is the block size. 
%If the number of neurons is significantly larger than $B$, then the overhead of loading $B$ can become prominent as we need to do multiple loads of $Y_l$ (Equation~\eqref{eq:activation}) to the shared memory. 
%That may become an overhead with large number of neurons since $B$ can be at most 1K. 
%To address this, we leverage thread coarsening in the optimized dense $\times$ sparse kernel, where each thread computes the multiple outputs of $C$ neurons, where $C$ is the coarsening factor. 
%Although thread coarsening increases the number output register \texttt{acc} and decreases the parallelism by $C\times$, but the net performance gain obtained can still be significant. This is because loading $Y_l$ to the shared memory is a costly operation compared to the effect of thread coarsening. 
%To illustrate thread coarsening we use $C=2$ in Figure~\ref{fig:optimized_denseXsparse} and $t_0$ computes outputs of two neurons in the same feature row.

\subsubsection{Efficient Access to Weight Matrix}
Even though the baseline CSR storage is efficient in terms of memory footprint, the access is not coalesced among threads in each warp. 
To address this, we store the weight matrix in a transposed  sliced-ELL storage format with zero-padding in warp granularity. 
Figure~\ref{fig:optimized_denseXsparse} shows the \texttt{wdispl} and \texttt{windex} data structures corresponding to the CSR representation shown in Figure~\ref{fig:naive_denseXsparse}.

The first two columns of \texttt{windex} (in orange) in Figure~\ref{fig:optimized_denseXsparse}(b) shows the layout of the elements for $block_{0,y}$. The top two sections separated by the dashed line each contains the \texttt{windex} elements accessed by a warp of the block during the first stage of execution. The bottom two sections are accessed by the same two warps during their second stage of execution.
The \texttt{wdispl} elements marks the positions of the dash lines and solid lines for every warp and every thread block.

The padded zeros are highlighted with red font color. The dashed lines represents the boundary between warps (each block has two warps in this example) and solid lines represent the boundaries between buffer stages, i.e., all blocks except block 2 involve two stagings and block 3 involves three stagings. In this example, the zero padding overhead is 27.5\% for warp granularity, however, it would be 80\% and 100\% with zero padding in tile and layer granularity, respectively.
Warp-level padding introduces a small number of zeros while maintaining coalesced (efficient) memory access.

\subsection{Memory Optimizations}

\subsubsection{Out-of-Core Storage and Overlapping Strategy}
\label{sec:overlap}
Data parallelism provides good scaling in an embarrassingly parallel fashion. However, duplicating weights in each GPU can make large networks infeasible for GPUs with limited memory capacity.
As a remedy, we implement an out-of-core storage algorithm that loads the required weight data structures to GPU memory from CPU memory when needed. Even though the out-of-core algorithm saves significant amount of memory, it has a data transfer overhead. 
We address this overhead by hiding the data transfer behind the GPU kernel with a double-buffering and overlapping strategy.

Double-buffering involves a pair of buffers in the GPU memory.
While layer $l$ uses weights from one buffer, weights for layer $l+1$ are moved into the other buffer.
When layer $l$ and the copy are both finished, the buffer pointers are swapped and the same procedure is followed for the next layer.
The data transfers are completely hidden behind the inference computation in our implementation.

\subsubsection{Compact Representation and Batching}
In order to reduce the memory footprint, we store the \texttt{map} and \texttt{windex} data structures with two-byte \texttt{unsigned short} data type. That reduces the memory footprint (and hence the data transfer time) by approximately 33\%.

We create batches for inputs to further reduce the memory consumed by input and output features during inference computation.
% We overcome another memory capacity limitation for storing the features, we batch \RN{by batching?} the input processing, i.e., each GPU first processes a batch of inputs and then another batch in a serial manner. 
Batching has no significant overhead since we overlap transfer of weights during the GPU computation as discussed in $\S$~\ref{sec:overlap}. 
As a result, we can fit even the largest inference problem in a single V100 GPU with 16 GB memory.

\section{Evaluation}
\label{sec:results}
\subsection{Experimental Setup}
% This paper employs a set of synthetic sparse neural networks as a testbed for the inference performance of our implementation. Even though Kepner \textit{et al}., provides the fine details and production of these synthetic networks~\cite{kepner2019}, we herein provide a brief overview of the provided dataset by the sparse deep neural network graph challenge.

  %22-core POWER9 & RHEL 7.6 & 4.14.0-115.21.2.el7a.ppc64le & V100-SXM2-16GB & 418.116 & Spectrum 10.3.1.2 & 10.1.243 & xl 16.1.1 \\

We use the Summit~\cite{summit} system at Oak Ridge National Laboratory.
Summit comprises 4,608 compute nodes, each with six 16 GB V100 GPUs.
%Triplets of GPUs are associated with each CPU; GPUs 0-2 with socket 0 and GPUs 3-5 with socket 1.
%Within a triplet, components are fully-connected by NVLink 2.0 x2 links, for 100 GB/s bidirectional bandwidth.
%Between triplets, the sockets are connected with a 64 GB/s X-bus SMP interconnect.
The network is a non-blocking fat tree of EDR InfiniBand with 23 GB/s node injection bandwidth.
% Each V100 supports mixed 16- and 32-bit floating point operations, which we use to accelerate our fastest implementation.
% Our implementation overcomes the 16GB memory of the V100, which is too small to fit the largest Sparse DNN Challenge networks.
The throughput numbers are calculated by calculating input edges over inference time. Inference time includes overlapped data copy time of weight and inputs to the GPUs.
Our evaluation is carried out using Spectrum MPI 10.3.1.2, XL compiler 16.1.1, nvcc 10.1.234, and CUDA driver 418.116;

\begin{table*}[]
\centering
% \captionsetup{font=large}
\vspace{1cm}
\caption{Performance (Edges/Second) and Speedup ($\times$) Comparisons with 2019 Sparse DNN Challenge Submissions}
\label{tab:comparison}
\renewcommand{\arraystretch}{1.5}
\resizebox{\textwidth}{!}{%
\begin{tabular}{ccccccccccccc}
 &  & \textbf{This Work} & \multicolumn{2}{c}{\textbf{Bisson \& Fatica}\cite{bisson2019gpu}} & \multicolumn{2}{c}{\textbf{Davis et al.}\cite{davis2019write}} & \multicolumn{2}{c}{\textbf{Ellis \& Rajamanickam}\cite{ellis2019scalable}} & \multicolumn{2}{c}{\textbf{Wang et al.}\cite{wang2019accelerating}} & \multicolumn{2}{c}{\textbf{Wang et al.}\cite{wang2019performance}} \\
 &  &  & \multicolumn{2}{c}{\textbf{2019 Champion}} & \multicolumn{2}{c}{\textbf{2019 Champion}} & \multicolumn{2}{c}{\textbf{2019 Innovation}} & \multicolumn{2}{c}{\textbf{2019 Student Innov.}} & \multicolumn{2}{c}{\textbf{2019 Finalist}} \\
\textbf{Neurons} & \textbf{Layers} & \textbf{Throughput} & \textbf{Throughput} & \textbf{Speedup} & \textbf{Throughput} & \textbf{Speedup} & \textbf{Throughput} & \textbf{Speedup} & \textbf{Throughput} & \textbf{Speedup} & \textbf{Throughput} & \textbf{Speedup} \\ \Xhline{0.75pt}
\multirow{3}{*}{\textbf{1024}} & \textbf{120} & 2.917E+13 & 4.517E+12 & 6.46 & 1.533E+11 & 190.28 & 2.760E+11 & 105.69 & 1.407E+11 & 207.32 & 8.434E+10 & 345.88 \\ \cline{2-13} 
 & \textbf{480} & 2.930E+13 & 7.703E+12 & 3.80 & 2.935E+11 & 99.83 & 2.800E+11 & 104.64 & 1.781E+11 & 164.51 & 9.643E+10 & 303.84 \\ \cline{2-13} 
 & \textbf{1920} & 2.883E+13 & 8.878E+12 & 3.25 & 2.754E+11 & 104.68 & 2.800E+11 & 102.96 & 1.896E+11 & 152.06 & 9.600E+10 & 300.30 \\  \Xhline{0.75pt}
\multirow{3}{*}{\textbf{4096}} & \textbf{120} & 8.220E+13 & 6.541E+12 & 12.57 & 1.388E+11 & 592.22 & 2.120E+11 & 387.74 & 1.943E+11 & 423.06 & 6.506E+10 & 1,263.52 \\ \cline{2-13} 
 & \textbf{480} & 8.222E+13 & 1.231E+13 & 6.68 & 1.743E+11 & 471.72 & 2.160E+11 & 380.65 & 2.141E+11 & 384.03 & 6.679E+10 & 1,230.99 \\ \cline{2-13} 
 & \textbf{1920} & 8.232E+13 & 1.483E+13 & 5.55 & 1.863E+11 & 441.87 & 2.160E+11 & 381.11 & 2.197E+11 & 374.69 & 6.617E+10 & 1,244.02 \\  \Xhline{0.75pt}
\multirow{3}{*}{\textbf{16384}} & \textbf{120} & 1.469E+14 & 1.008E+13 & 14.57 & 1.048E+11 & 1,401.53 & 1.270E+11 & 1,156.54 & 1.966E+11 & 747.10 & 3.797E+10 & 3,867.84 \\ \cline{2-13} 
 & \textbf{480} & 1.394E+14 & 1.500E+13 & 9.29 & 1.156E+11 & 1,206.23 & 1.280E+11 & 1,089.38 & 2.060E+11 & 676.89 & 3.747E+10 & 3,721.66 \\ \cline{2-13} 
 & \textbf{1920} & 1.464E+14 & 1.670E+13 & 8.77 & 1.203E+11 & 1,216.96 & 1.310E+11 & 1,117.56 & 1.964E+11 & 745.52 & 3.750E+10 & 3,903.72 \\  \Xhline{0.75pt}
\multirow{3}{*}{\textbf{65536}} & \textbf{120} & 1.796E+14 & 9.388E+12 & 19.13 & 1.050E+11 & 1710.29 & 9.110E+10 & 1971.24 & 1.892E11 & 949.15 & -- & -- \\ \cline{2-13} 
 & \textbf{480} & 1.703E+14 & 1.638E+13 & 10.40 & 1.091E+11 & 1,560.59 & 8.580E+10 & 1,984.38 & 1.799E+11 & 946.41 & -- & -- \\ \cline{2-13} 
 & \textbf{1920} & 1.714E+14 & 1.787E+13 & 9.59 & 1.127E+11 & 1,520.59 & 8.430E+10 & 2,032.86 & -- & -- & -- & -- \\  \Xhline{0.75pt}
\end{tabular}
}
\vspace{-4mm}
\end{table*}

\subsection{Single GPU performance}
\subsubsection{Single Volta V100 GPU performance}
The first column of Table~\ref{tab:summit} shows the single-GPU throughput of our optimized implementation.
Compared to the baseline implementation, our optimizations provide 5.56$\times$--11.84$\times$ performance improvement. Thus, we do not report the baseline performance in detail due to lack of space. 
As the depth of the network increases, the throughput increases due to increased average sparsity in the features.
This means more feature rows are entirely zero, and no thread-block is mapped to them.
As the number of neurons increases, the throughput decreases due to two effects.
First, increased zero-padding in the sliced ELLPACK format causes more wasted work and memory bandwidth.
Second, less reuse from shared memory, since a given set of outputs is less likely to reuse the same footprint of features.

\subsubsection{Single Ampere A100 GPU Performance}
We now compare the performance of the optimized fused kernel on the latest GPU, NVIDIA Ampere A100.
% NVIDIA A100 GPU is state-of-the art accelerator for DNN computation~\cite{amperewhite}. %Vikram: I am good with this removal. :D
A100 increases memory capacity from 16 to 40 GB, grows L2 cache from 6MB to 40MB, and brings 1.73$\times$ global memory bandwidth and 1.24$\times$ the single-precision floating point peak performance~\cite{amperewhite}. 
At the time of writing, the A100 is not available to the public, so we do not have a cluster for scaling studies.
We restrict our study to the performance improvement of our optimized-fused kernel on out-of-the box single A100 GPU, without any A100-specific optimizations or tuning.
We use CUDA 11 with NVIDIA driver 450.51 on an Ubuntu machine. 

Table~\ref{tab:summit} shows the performance improvement achieved by the our optimized implementation on Ampere A100 GPU.
Using the same code, A100 yields a 1.45$\times$ to 2.37$\times$ speedup.
The improvement is more modest for smaller networks, for which the implementation makes only modest demand on the memory subsystem.
For larger networks where the kernel relies more on cache and global memory performance, A100 provides a much larger speedup.
Optimization specific to A100's architectural features should yield further improvement.

\subsection{Multi-GPU Scaling on Summit System}
In this evaluation, we report scaling of our sparse DNN inference code with the Sparse DNN Challenge dataset.
We use MPI to parallelize the inference across multiple GPUs on the Summit system without any network architecture- or topology-specific optimization.
This work employs a batch parallelism strategy with an embarrassingly-parallel fashion, where weights are replicated between GPUs and the features are partitioned evenly across GPUs. During the inference, we prune the inactive features, which yields load imbalance between GPUs. Also, when the number of remaining active features per GPU is small, the per-GPU throughput drops significantly. We leave these open problems for future work.

Table~\ref{tab:summit} shows the throughput achieved on Summit for all network configurations on up to 768 GPUs.
For the smallest network configuration, the strong-scaling limit is 16 nodes (96 GPUs), which is more than the four-GPU scaling limit observed on a \textit{single} node of the 2019 champions.
Although strong scaling is observed, the largest parallel efficiency observed is 54\% at three nodes.
For larger configuration, strong scaling is observed out to 128 nodes (768 GPUs) and corresponds to 223.3 GigaEdges/Second per GPU, and a parallel efficiency of 87.6\% is observed for one full node (six GPUs), and 82.6\% at two nodes.
% \vikram{For large networks, the load-balancing helps maintain some scaling out to 768 nodes, while for smaller networks there is not enough work to justify the load-balancing cost, and parallel efficiency suffers.}

\subsection{Comparison with Prior Work}
Table~\ref{tab:comparison} compares the fastest time from our submission with the fastest times from various 2019 submissions.
% ``Througput'' is the reported edges-per-second, and ``sp'' is our submission's speedup over their throughput.
Our speedup varies from 3.25$\times$ to 19.13$\times$ over the fastest champion from 2019, with one configuration achieving a 1710$\times$ speedup over one of the 2019 champions.
Part of the contribution is from the kernel implementation: the single-GPU speedup over Bisson \& Fatica~\cite{bisson2019gpu} varies from 4.3$\times$ for the 1024-neuron 120-layer to 1.13$\times$ for the 65526-neuron 1920-layer configurations.

\subsubsection{Comparison with cuSPARSE Library}
Wang et al.~\cite{wang2019performance} uses cuSPARSE on a V100 in their challenge submission in 2019.
Like this work, they do not operate on inactive features, making it possible for a direct comparison.
As they report single-GPU inference (the last column of Table~\ref{tab:comparison}) times on V100, we are able to compare our single-GPU implementation to the fastest results they achieved with cuSPARSE.
The speedup of our single-GPU implementation varies from 210$\times$ for 4096 neurons / 1920 layers to 125$\times$ for 1024 neurons / 120 layers over their cuSPARSE implementation.

\section{Conclusion}
\label{sec:conclusion}

In this work, we discuss that a baseline sparse DNN kernel performance is limited by memory bandwidth and irregular memory accesses. 
To address this, we propose three performance optimizations: register tiling, shared-memory tiling, and efficient memory access to weight matrices. We scale our implementation up to 768 Volta V100 GPUs and report up to 180 TeraEdges/Second sustained inference throughput. 
These results are up to 4.3$\times$ faster on a single V100 GPU and an order of magnitude faster at scale when compared with 2019 Champions. 
We also show that proposed implementation when executed with the latest Ampere A100 GPU, without any optimization, achieves up to 2.37$\times$ speedup over V100 GPU. 

\section*{Acknowledgment}
The authors acknowledge Kishore Iyer, Jingning Tang, Hanhaotian Liu, and Volodymyr Kindratenko for their help. 
This research used resources of the Oak Ridge Leadership Computing Facility at the Oak Ridge National Laboratory, which is supported by the Office of Science of the U.S. Department of Energy under Contract No. DE-AC05-00OR22725 and also utilized resources supported by the National Science Foundation’s Major Research Instrumentation program, grant \#1725729, as well as the University of Illinois at Urbana-Champaign. 
This work is supported by IBM-ILLINOIS Center for Cognitive Computing Systems Research (C3SR) and
partly supported by the Center for Applications Driving Architectures (ADA) and Center for Research on Intelligent Storage and Processing-in-memory (CRISP), JUMP Centers with prime award coming from SRC.

\iffalse
\section*{Acknowledgment}

This material was partially supported by the U.S. Department of Energy, Office of Science, Advanced Scientific Computing Research and Basic Energy Sciences, under Contract DE-AC02-06CH11357. 
We gratefully acknowledge the computing resources provided and operated by the Oak Ridge Leadership Computing Facility, which is a U.S. Department of Energy, Office of Science User Facility. 
We thank Narayanan (Bobby) Kasthuri from UChicago/Argonne, and Rafael Vescovi and Ming Du from Argonne for sharing the mouse brain dataset. 
The mouse brain, IC chip and activated charcoal data were collected at 32-ID beamline by Vincent De Andrade at Advanced Photon Source, Argonne National Laboratory.
This research is part of the Blue Waters sustained-petascale computing project, which is supported by the National Science Foundation (awards OCI-0725070 and ACI-1238993) and the state of Illinois. 
Blue Waters is a joint effort of the University of Illinois at Urbana-Champaign and its National Center for Supercomputing Applications. 
This research is based in part upon work supported by: The Center for Applications Driving Architectures (ADA), a JUMP Center co-sponsored by SRC and DARPA. 
This material is based in part upon work supported by the XPACC Center for Exascale Simulation of Plasma-Coupled Combustion and Department of Energy, under Award Number DE-NA0002374.
\fi

%\balance
%% Bibliography
\bibliographystyle{ieeetr}
\bibliography{main}

\end{document}